\documentclass{iau} 
\usepackage{graphicx}
\usepackage{comment}
\usepackage{natbib}

\newcommand{\bsy}{\boldsymbol}
\newcommand{\delm}{\delta{}m}

\title[Solar-System Studies with PTAs] 
{Solar-System Studies \\ with Pulsar Timing Arrays}

\author[R.~N.~Caballero and Collaborators]   
{R.~N.~Caballero$^1$ and Collaborators}
\affiliation{$^1$Max-Planck-Institut f{\"u}r Radioastronomie, Auf dem H{\"u}gel 69, 53121 Bonn, Germany \\ email: {\tt caball@mpifr-bonn.mpg.de} \\[\affilskip]
}

\pubyear{2017}
\volume{337}  
\jname{Pulsar Astrophysics -­ The Next 50 Years}
\editors{P. Weltevrede, B.B.P. Perera, L. Levin Preston \& S. Sanidas, eds.}

\begin{document}

\maketitle

\begin{abstract}
High-precision pulsar timing is central to a wide range of astrophysics and fundamental physics applications.
When timing an ensemble of millisecond pulsars in different sky positions, known as a pulsar
timing array (PTA), one can search for ultra-low-frequency gravitational waves (GWs) through
the spatial correlations that spacetime deformations by passing GWs are predicted to induce on the
pulses' times-of-arrival (TOAs). A pulsar-timing model, requires the use of a solar-system
ephemeris (SSE) to properly predict the position of the solar-system barycentre, the (quasi-)inertial
frame where all TOAs are referred. Here, I discuss how while errors in SSEs can introduce
correlations in the TOAs that may interfere with GW searches, one can make use of PTAs to study
the solar system. I discuss work done within the context of the European Pulsar
Timing Array and the International Pulsar Timing Array collaborations. These include new updates
on the masses of planets from PTA data, first limits on masses of the most massive asteroids,
and comparisons between SSEs from independent groups. Finally, I discuss a new
approach in setting limits on the masses of 
unknown bodies in the solar system and calculate mass sensitivity curves for PTA data.
\end{abstract}

\firstsection 
\section{Introduction}

A pulsar timing array (PTA), is an ensemble of pulsars in various sky positions,
and can be employed to search for processes that cause
space-correlated signals which will be present in all pulsars  \cite[e.g.][]{fb1990}. 
PTA research is based on
precision pulsar timing, i.e. the modelling of pulse times-of-arrival (TOAs),
recorded at high precision \cite[e.g.][]{lk2005}. 
The signals of interest are sought in the timing residuals, i.e.
the differences between the observed and model-predicted TOAs, which 
means that the sensitivity of PTAs to such effects is limited by the rotational stability
of the pulsars, which work as celestial clocks. Therefore, PTAs are constructed
using the most rotationally stable pulsars known, the millisecond pulsars (MSPs).

PTAs focus primarily on efforts for the direct detection of gravitational waves (GWs)
in the frequency regime $10^{-9}-10^{-6}$~Hz \cite[e.g.][]{sv2010}. 
Three collaborations are 
actively working in the realisation of this goal. Citing their latest data releases, these are
the European Pulsar Timing Array \cite[EPTA;][]{dcl2016} in Europe,
the Parkes Pulsar Timing Array \cite[PPTA;][]{rhc2016} in Australia,
and the North-American Nanohertz Observatory for 
Gravitational Waves  \cite[NANOGrav;][]{abb2015} in North America.
These collaborations work together under the
International Pulsar Timing Array consortium \cite[IPTA;][]{vlh2016}, 
in an effort to improve both the sensitivity and the robustness of the data
and data analyses.

PTA sensitivity to GWs is particularly limited by
possible spatially-correlated signals other than GWs \citep[e.g.][]{thk2016},
particularly in the lower-to-intermediate signal-to-noise regime.
The two most frequently studied sources of spatially-correlated noise
are possible errors in the terrestrial time-standards to which all TOAs are referred
and in the solar-system ephemeris (SSE) used. While PTAs are actively
working into mitigating these sources of noise, their understanding actually 
allows PTAs to simultaneously expand their scientific studies.
In relation with these noise components specifically, PTA data
are used to construct a pulsar-based time-scale of such long-term
stability that it can serve as an independent cross-check for the highly
precise time-scales from atomic clocks \cite{hcm2012}, and to provide
constraints on possible errors in SSEs and the 
masses of solar-system planets \cite[][henceforth CHM10]{chm2010}.
Here, we focus on the latter application and discuss work conducted 
within the framework of the regional PTAs and the IPTA. 

\section{Solar-system ephemerides and Pulsar Timing}

One of the basic reasons that pulsar-timing models use SSEs,
is to predict the position of the solar-system barycentre (SSB)
for every observing epoch. This is essential, since before 
calculating the pulse-emission time at the pulsar's co-moving frame,
we refer all TOAs to the SSB, a common (quasi-)inertial reference frame.
Using PPTA data, CHM10 were the first to use pulsar timing to constrain
the masses of planets. The developed method was focused on 
the approximation of small errors in the planetary masses
used by the SSE,
i.e. when the error is much smaller than the total mass of the planet. 
In such an approximation, the effect of 
the small error in mass, $\delm$,
is a small displacement of the SSB along 
barycentric position vector of the planet, 
with respect to the original SSB position.
One can show how this delay
depends on the pulsar position, by expressing the
timing signature of such an effect in terms of the
differences in the ecliptic latitude and longitude 
between the (i-th) planet and the (j-th)
pulsar, $\Delta\beta_{i,j}$ and $\Delta\lambda_{i,j}$ respectively, as
\begin{equation}
\label{eq:dmass2}
\tau_{b} \propto |\bsy{b}|\delm{}_{i} \cos(\Delta\beta_{i,j})\cos(\Delta\lambda_{i,j})\,,
\end{equation} 
where $|\bsy{b}|$ is the determinant of the barycentric position vector of the planet. 
It is apparent, that 
fitting multiple pulsars located in as many different sky position as possible
with $\delm{}$ as a global fit parameter, 
improves the quality of the $\delm{}$ estimation. 

Using four MSPs, CHM10 found that the
possible errors in the masses where consistent
with zero at the 2$\sigma$ level. The data length
of the pulsars ranged from 5.2 to 22.1~yr, 
which meant that for the giant planets with
longer periods, the measurements should improve 
significantly once more MSPs contribute with time-spans 
longer than the orbital period of Jupiter (11.86~yr).
Finally, CHM10 used only one specific SSE.  
Knowing that there are differences between
SSEs, as discussed below, an interesting idea has been to
see whether such differences will affect the
measured planetary masses by PTAs.

\section{Current and future work}

Various groups are creating SSEs, primarily
used for space-mission navigation.
In return, in situ measurements of planetary 
masses by spacecrafts
such as the \emph{Pioneer} and the \emph{Voyager}
provide input for future SSEs. So far,
pulsar timing have primarily been using 
SSEs from two independent groups, namely
the DE and the INPOP series of SSEs from the Jet Propulsion Laboratory (JPL)
and the Institut de M\'ecanique C\'eleste et de Calcul des \'Eph\'em\'erides (IMCCE),
respectively.
These ephemerides are created using a wealth of data
from sources such as optical astrometry, spacecraft mass measurements,
radar and laser ranging, etc (see CHM10 and references therein),
which are used as input in numerical integrations of the planetary equations of motion.

\begin{figure}
\begin{center}$
\begin{array}{cc}
\includegraphics[width=7.5cm, angle=0]{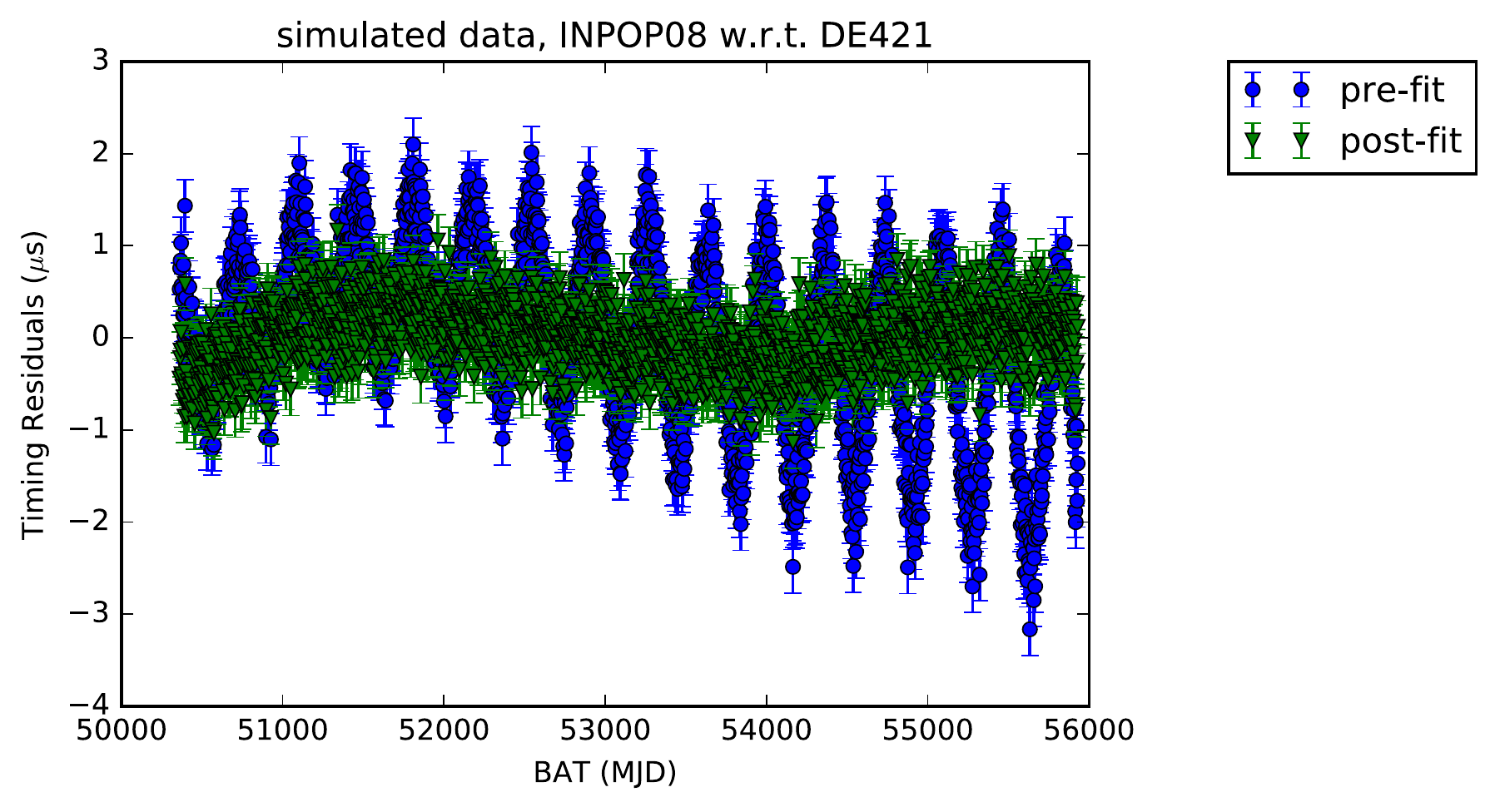} &
\includegraphics[width=6.5cm, angle=0]{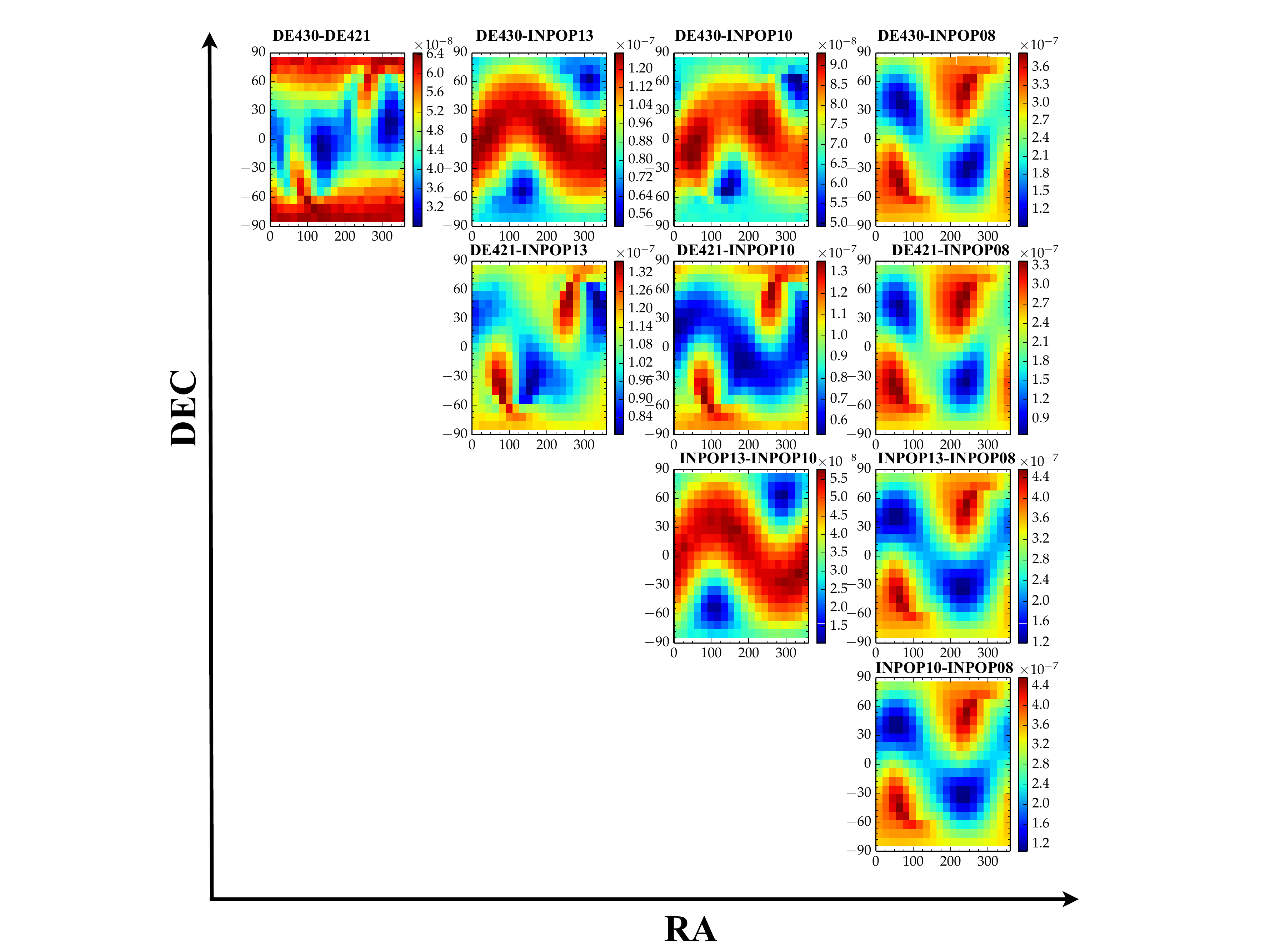} \\
\end{array}$
\caption{\emph{Left Panel}: Timing residuals of the 
difference between the SSEs DE421 and INPOP08, based on
simulated data. The figure emphasises the effects of fitting for the
pulsar timing model. The blue circles correspond to the difference before
fitting the pulsar modes (pre-fit) and have a larger root-mean-square than
the residuals after fitting the pulsar model (post-fit), denoted with the green triangles. 
\emph{Right Panel}: Difference in the timing residuals for various
pairs of SSEs, as a function of sky position, expressed in Right Ascension (RA)
and Declination (DEC). The colour-code (for colour prints) is the root-mean-square of the residuals.
For each case, we used simulations of 400 MSPs isotropically distributed in the sky. Note
how for some pairs the difference is stronger along the ecliptic plane, while for other
pairs, away from it.} 
\label{fig1}
\end{center} 
\end{figure} 

JPL and IMCCE regularly present differences between the various published SSEs.
In pulsar timing, it is important to
note that the timing-model fit will absorb some of
these differences, biasing some of the parameters.
For example, annual, linear and quadratic
variations are absorbed by fitting of the pulsar position,
pulsar rotational spin and spin-down, respectively.
In Fig.~\ref{fig1} (left panel), we see such a difference for simulated data
of PSR J1713+0747, with TOA uncertainties reflecting those
in the published EPTA data. To compare the SSEs, we simulate data using
one SSE and then fit the timing model using another SSE.
In order to understand the effects of using different 
SSEs in EPTA work, we examined the residuals of the difference between 
SSEs in many sky positions. Fig.~\ref{fig1} (right panel) shows the
the root-mean-square of the post-fit timing residuals
for the difference of various pairs of SSEs. For each case,
we used 400 MSPs isotropically distributed in the sky. 
In order to make the effects prominent we simulated the 
data with a precision of only a few ns. 
One can see, that as our signal-to-noise improves,
if PTAs are limited by the number of MSPs
contributing to the measurement of correlated effects, using different SSEs
will affect our sensitivity to the errors in planet masses.

Within the context of the IPTA, we are using data from the first
official data release \cite{vlh2016} to extend on the work of CHM10.
The IPTA data set constitutes a significant improvement
thanks to increased data-span and timing precision
for multiple of the most rotationally stable MSPs, 
better observing-frequency coverage,
increased number of MSPs contributing to the solutions,
and more sophisticated
methods for modelling the noise of individual pulsars \cite[e.g.][]{lsc2016}. 
Proper noise models are of central importance. Indeed, CHM10 excluded 
their most precisely-timed MSP from the mass measurement of Mars due to insufficiencies in
the noise model.
Preliminary results with the IPTA, indicate improvements of factors
$4-10$ by comparison to CHM10, depending on the planet. By comparison to
only using EPTA data, the IPTA data set, which combines data from all three PTAs,
gives an additional improvement of up to a factor $\sim{}4$. 
Furthermore, the IPTA data are now sensitive to mass errors in large
objects in the asteroid belt, with initial results constraining the mass of Ceres 
with precision only an order of magnitude below that published by the IAU \cite{lcf2011}. 
Our IPTA work also repeats the analysis for multiple SSEs from both the JPL and IMCCE.
Initial results that compare DE421, DE430, DE435 and INPOP06C showed
results that are statistically consistent. Unsurprisingly, deviations start becoming 
more prominent for the giant planets. This is due to the limited number of orbits
completed within the data time-span (if any) and to the higher degree of 
correlation between the timing signals and the pulsar's 
low-frequency noise components,
and the rotational spin period and period derivative. 
 
As our data become more sensitive, errors in masses that we may measure for the
giant planets could be due to errors in the mass determination of their satellites,
since PTAs are in fact sensitive to the planetary systems, 
rather than the planets or their satellites individually. 
For example, in the case of DE421, five Jovian and nine
Saturnian satellites were used in the model (CHM10). As such, dynamical modelling
of solar-system bodies in PTAs may allow to shed light in the origin
of possible mass-error measurements.
As a first step, we use a dynamical model to calculate the
effects on the pulsar residuals from unknown bodies in Keplerian orbits
around the SSB, 
essentially accounting only for the gravitational effects of the Sun 
(Guo, Lee \& Caballero; submitted), 
and estimate the IPTA-data 
sensitivity to such objects.
We note that this analysis is valid for any type of object, even hypothetical, 
such as dark-matter clumps \cite{lz1997}.
With increased precision, time-span and number of pulsars, it is possible that the IPTA
will eventually become sensitive to masses of the order of official IAU uncertainties for the Jovian system.
In the future, we would focus on expanding the dynamical model to include higher-order
effects, such as perturbations from objects other than the Sun, allowing PTAs to also study
the effects of satellites in orbit with giant planets and contribute more significantly to the development
of SSEs. Our results on constraining the masses of planets and unknown bodies in the 
solar system using the IPTA will be published 
in an IPTA paper (Caballero et al. IPTA Collaboration; in prep.).
  
\section{Acknowledgments}

The ongoing work discussed here is conducted in the framework of the EPTA
and the IPTA, and all members who contributed through
instrument development, observations, data analysis and algorithm development
are acknowledged. Results discussed in this contribution are based on data collected with 
Effelsberg Radio Telescope, the Nan\c cay Radio Telescope, the Lovell Telescope, the 
Westerbork Radio Synthesis Telescope, the Parkes Radio Telescope, the Green Bank Telescope,
and the Arecibo Radio Telescope.


\begin{thebibliography}{}  

\bibitem[Arzoumanian \etal(2015)]{abb2015}
{{Arzoumanian}, Z. et al. NANOGrav Collaboration} 2015,
\textit{ApJ}, 813, 65

\bibitem[Champion \etal(2010)]{chm2010}
{Champion, D.~J., Hobbs, G.~B., Manchester, R.~N. et al.} 1997,
\textit{ApJ}, 720, L201

\bibitem[Desvignes \etal(2016)]{dcl2016}
{Desvignes, G., Caballero, R.~N., Lentati, L. et al.} 2016,
\textit{MNRAS}, 458, 3341

\bibitem[Foster \& Backer(1990)]{fb1990}
{Foster, R.~S. \& Backer, D.~C.} 1990,
\textit{ApJ}, 361, 300

\bibitem[Hobbs \etal(2012)]{hcm2012}
{Hobbs, G., Coles, W., Manchester R.~N. et al.} 2012,
\textit{MNRAS}, 427, 2780

\bibitem[Lentati \etal(2016)]{lsc2016}
{Lentati L., Shannon, R.~M., Coles, W.~A.} 2016,
\textit{MNRAS}, 458, 2161

\bibitem[Loeb \& Zaldarriaga(1997)]{lz1997}
{Loeb, A. \& Zaldarriaga, M.} 1997,
\textit{Phys. Rev. D}, 71, 103520 

\bibitem[Lorimer \& Kramer(2005)]{lk2005}
{Lorimer, D.~R. \& Kramer, M.} 2005,
\textit{Handbook of Pulsar Astronomy. Cambridge Univ. Press}

\bibitem[Luzum \etal(2011)]{lcf2011}
{Luzum, B., Capitaine, N., Fienga, A. et al.} 2011,
\textit{Cel. Mech. \& Dynamical Astron.}, 110, 293, 

\bibitem[Reardon \etal(2016)]{rhc2016}
{Reardon, D.~J., Hobbs, G., Coles, W. et al.} 2016,
\textit{MNRAS}, 455, 1751 

\bibitem[Sesana \& Vecchio(2010)]{sv2010}
{Sesana, A. \& Vecchio, A.} 2010,
\textit{Phys. Rev. D}, 81, 104008 

\bibitem[Tiburzi \etal(2016)]{thk2016}
{Tiburzi, C., Hobbs, G., Kerr} 2016,
\textit{MNRAS}, 455, 4339

\bibitem[Verbiest \etal(2016)]{vlh2016}
{Verbiest, J.~P.~W., Lentati, L., Hobbs, G. et al.} 2016,
\textit{MNRAS}, 458, 1267

\end{thebibliography}
\end{document}